\documentstyle[preprint,aps]{revtex}
\input epsf.tex
\def\DESepsf(#1 width #2){\epsfxsize=#2 \epsfbox{#1}}
\begin{document}
\preprint{\vbox{\hbox{}}}
\draft
\title{Contribution to $\epsilon'/\epsilon$ from anomalous gauge couplings
}
\author{
Xiao-Gang He
}

\address{
Department of Physics, National Taiwan University, Taipei, 10764
}

\date{March 1999}
\maketitle
\begin{abstract}
Recently KTev collaboration has measured $Re(\epsilon'/\epsilon) = (28\pm 4.1)
\times 10^{-4}$ which is in agreement with early measurement from NA31. 
The Standard Model prediction for $\epsilon'/\epsilon$ is on the lower end 
of the experimentally allowed range depending on models for hadronic matrix 
elements. In this paper we study 
the contributions from 
anomalous gauge couplings. We find that the contributions 
from anomalous couplings can be significant and can ehance $\epsilon'/\epsilon$ to have 
a value closer to data.
\end{abstract}
\pacs{PACS numbers: }

The parameter $\epsilon'/\epsilon$ measuring direct 
CP violation in $K\to \pi\pi$ is a very important quantity to study\cite{review}. A non-zero value 
$Re(\epsilon'/\epsilon) = (2.3\pm 0.65)\times 10^{-3}$ was first measured by 
NA31 experiment\cite{1}, but E731 experiment with similar sensitivity did not confirm it\cite{2}. This 
controverse is now settled with the measurement of $Re(\epsilon'/\epsilon) 
=(2.8\pm 0.41)\times 10^{-3}$ by KTev experiment\cite{3}. The parameter $\epsilon'/\epsilon$
 is different from
 the parameter $\epsilon$ which characterizes
CP violation in $K^0-\bar K^0$ mixing. CP violation due to mixing was first observed in 1964. 
Before the measurement of $\epsilon'/\epsilon$ 
the non-zero value\cite{4} $\epsilon =2.266\times e^{i\phi_\epsilon}$ with 
$\phi_\epsilon \approx \pi/4$ 
is the only laboratory experimental evidence for CP violation.
Many models have been proposed to explain the non-zero value for $\epsilon$\cite{5}. 
One class of models
 is the superweak models\cite{6} of CP violation which postulate that there is a new 
$\Delta S = 2$ interaction causing mixing and CP violation in $K^0-\bar K^0$ system.
In such models, 
there is no CP violation in $\Delta S = 1$ interaction and $\epsilon'/\epsilon$ is predicted 
to be zero.
 Therefore the confirmation of non-zero
$\epsilon'/\epsilon$, now, decisively rules out superweak models. 

There are other models, such as the standard Kobayashi-Maskawa model (SM)\cite{7}, 
and the multi-Higgs model\cite{8}, 
which not only violate CP in $\Delta S =2$, but also in $\Delta S = 1$
interactions.
These non-superweak 
models
with different CP violating mechanisms  can explain the 
measured $\epsilon$ and predict, in general, different values for 
$\epsilon'/\epsilon$. It is clear that $\epsilon'/\epsilon$ 
can provide further information about models for CP violation.
In the SM the predicted value for $\epsilon'/\epsilon$, although in the lower end allowed
by experiment as will be seen later, is not in conflict with data.
More studies on the uncertainties associated with the relevant 
hadronic matrix elements are needed.
There is also the possibility that new physics does contribute and alter the 
SM prediction significantly. 
In this paper we reanalyse the 
anomalous gauge coupling effects on $\epsilon'/\epsilon$ in a similar way as in Ref.\cite{9}.
We find that the anomalous couplings can affect $\epsilon'/\epsilon$ significantly. Using the 
measured value for $\epsilon'/\epsilon$, one can also constrain the allowed range 
for the anomalous couplings.

The CP violating $\Delta S = 1$ interaction 
responsible for $K\to \pi\pi$ decays
in the SM is dominated by the strong penguin contributions
which contribute to $I=0$ amplitude $A_0$ only. The contributions due to
electroweak penguins  which contribute to both 
$I=0$ and $I=2$ amplitudes $A_0$ and $A_2$, are small.
However the electroweak penguins contribute to $\epsilon'/\epsilon$ 
significantly because there is a well-known enhancement 
factor $1/\omega=ReA_0/Re A_2 =22.2$ for $I=2$ contributions\cite{10}.
For the same reason, although the contributions from anomalous 
couplings are similar in size to those from the electroweak penguins, these contributions
having both $I=0$ and $I=2$ components can thus 
affect $\epsilon'/\epsilon$ significantly.

The most general $WWV$ interactions with the $W$ gauge boson on-shell and invariant 
under $U(1)_{em}$ can be parametrized as\cite{11}

\begin{eqnarray}
L_V &=& -ig_V [\kappa^V W^+_\mu W^-_\nu V^{\mu\nu} +
{\lambda^V\over m_W^2} W^+_{\sigma\rho} W^{-\rho \delta} V_\delta^\sigma
\nonumber\\
&+&\tilde \kappa^V W^+_\mu W^-_\nu \tilde V^{\mu\nu}
+ {\tilde \lambda^V\over m^2_W} W^+_{\sigma\rho}W^{-\rho\delta}
\tilde V_\delta^\sigma\nonumber\\
&+&g_1^V(W^{+\mu\nu}W^-_{\mu\nu} - W^+_\mu W^{-\mu\nu})V_\nu
+g_4^V W^+_\mu W^-_\nu (\partial^\mu V^\nu + \partial^\nu V^\mu)
\nonumber\\
&+& g_5^V \epsilon_{\mu\nu\alpha\beta} (W^{+\mu}
\partial^\alpha W^{-\nu} - \partial^\alpha W^{+\mu}W^{-\nu}) V^\beta],
\end{eqnarray}
where $W^{\pm \mu}$ are the $W$ boson fields; $V$ can be the $\gamma$ or
$Z$ fields; $W_{\mu\nu}$ and $V_{\mu\nu}$ are the $W$ and $V$ field strengths,
respectively; and $\tilde W (\tilde V)_{\mu\nu} = \epsilon_{\mu\nu\alpha\beta}
W(V)^{\alpha\beta}/2$. The terms proportional to $\kappa$, $\lambda$, and
$g_{1,5}$ are CP conserving and $\tilde \kappa$, $\tilde \lambda$, and
$g^Z_4$ are CP violating. Our convention is that for $V = \gamma$, $g_V = e$ and for 
$V = Z$, $g_V = g\cos\theta_W$. $g^\gamma_1$ defines the $W$ boson charge and is
always normalized to 1. In the SM at the tree level, $\kappa^V$ and $g_1^Z$
are equal to 1, and all others are zero. We refer 
$\Delta \kappa^V = \kappa^V -1$, $\Delta g_1^Z -1$, $\tilde \kappa^V$,
$\tilde \lambda$, $g_4^V$ and $g_5^V$ as anomalous couplings.

With non-zero anomalous couplings, at the energy scale $\mu = m_W$ 
beside the SM effective Hamiltonian $H_{SM}$ for $\Delta S = 1$ interaction,
there are additional contributions\cite{9}

\begin{eqnarray}
H_{AC} &=&
{G_F\alpha_{em} \over 2 \sqrt{2} \pi} \sum_{i=u,c,t}V_{id}V_{is}^*
\sum_{q=u,d}[Q_q H(x_i)_A \bar s \gamma_\mu (1-\gamma_5) d \bar q \gamma^\mu q\nonumber\\
&+& \cot^2\theta_W F(x_i)_A \bar s \gamma_\mu (1-\gamma_5) d
\bar q \gamma^\mu (Q_q \sin^2\theta_W - T^3 {1-\gamma_5\over 2} )q],
\end{eqnarray}
where $Q_q$ is the charge of $q$ quark, $x_i = m_i^2/m_W^2$,
$T_3$ is the isospin operator with eigenvalues $1/2$ and $-1/2$ for $u$ and $d$ respectively,
and

\begin{eqnarray}
H(x)_A &=& \Delta\kappa^\gamma {x\over 4} \ln {\Lambda^2\over m^2_W}
+\lambda^\gamma [{x(1-3x)\over 2(1-x)^2}-{x^3\over (1-x)^3}\ln x],\nonumber\\
F(x)_A&=& -\Delta g_1^Z {3\over 2}x \ln {\Lambda^2\over m^2_W}
+g_5^Z [{3x\over 1-x} + {3x^2\over (1-x)^2}\ln x].
\end{eqnarray}
In obtaining $H_{AC}$ we have used unitary gauge 
and introduced a momentum cut-off $\Lambda$ for terms which
are divergent in loop integrals. Note that $H_A$ and $F_A$ are proportional to 
internal quark mass squared. The dominant contribution is 
therefore from t quark in the
loop. 
Also to the leading order $H_{AC}$ does not depend on 
CP violating couplings $\tilde \kappa^\gamma$, $\tilde \lambda^\gamma$ and $g_4^Z$. 
Contributions from these couplings and $\Delta \kappa^Z$ are suppressed by 
factors of $O((m^2_{d,s}, m_K^2)/m^2_W))$ and are 
neglected. The source of CP violation for $K\to \pi\pi$ decays 
with anomalous couplings is therefore the
same as in the SM.

At an  energy scale $\mu$ lower than $m_W$ the 
effective Hamiltonian $H_{eff}$ for $\Delta S = 1$ interaction 
receives important QCD corrections. $H_{eff}$ 
is usually 
parametrized as\cite{12}

\begin{eqnarray}
H_{eff}&=&
{G_F\over \sqrt{2}} V_{ud}V_{us}^*\sum_{i=1,10}
C_i(\mu) O_i(\mu),
\end{eqnarray}
where $C_i(\mu) = z_i(\mu) + \tau y_i(\mu)$, $\tau = -V_{td}V_{ts}^*
/V_{ud}V_{us}^*$, and

\begin{eqnarray}
&&O_{1}=\bar s \gamma_\mu (1-\gamma_5) d \bar u \gamma^\mu (1-\gamma_5) u, \;\; 
O_2 = \bar s \gamma_\mu (1-\gamma_5) u \bar u \gamma^\mu 
(1-\gamma_5)  d,\nonumber\\
&&O_{3,5} = \bar s \gamma_\mu (1-\gamma_5)d \sum_q \bar q \gamma^\mu (1\mp\gamma_5) q,
\;\;
O_{4,6}= \bar s_{\alpha} \gamma_\mu (1-\gamma_5) d_{\beta}
\sum_q \bar q_\beta \gamma^\mu (1\mp\gamma_5) q_\alpha,\nonumber\\
&&O_{7,9} = {3\over 2}\bar s \gamma_\mu (1-\gamma_5) d \sum_q Q_q
\bar q \gamma^\mu (1\pm\gamma_5)) q,
\;\;
O_{8,10}= {3\over 2}\bar s_{\alpha} \gamma_\mu (1-\gamma_5) d_{\beta}
\sum_q Q_q\bar q_\beta \gamma^\mu (1\pm\gamma_5) q_\alpha.
\end{eqnarray}
In the above we have neglected the dipole interactions which have been shown to have negligible
contributions to $\epsilon'/\epsilon$\cite{dhp}.

The boundary conditions of the Wilson coefficients (WC) 
needed for renormalization running from $m_W$ to 
$\mu$ for the SM can be found in Ref.\cite{12} and the 
ones due to anomalous couplings are

\begin{eqnarray}
y^{AC}_3(m_W) &=&-{\alpha_{em} \over 24\pi} F(x_t) \cot^2\theta\,\nonumber\\
y^{AC}_7(m_W)&=&-{\alpha_{em}\over 6\pi}[H(x_t)_A + \sin^2\theta_W \cot^2\theta_W 
F(x_t)_A]\,,
\nonumber\\
y^{AC}_9(m_W)&=& -{\alpha_{em}\over 6\pi} [H(x_t)_A - \cos^2\theta_W 
\cot^2\theta_W F(x_t)_A].
\end{eqnarray}

In the SM the WCs $y^{SM}_i$ have been obtained at the next-leading order in QCD. 
We will use
the values,

\begin{eqnarray}
&&(y^{SM}_3, y^{SM}_4, y^{SM}_5,y^{SM}_6, y^{SM}_7/\alpha_{em}, y^{SM}_8/\alpha_{em}, 
y^{SM}_9/\alpha_{em}, y^{SM}_{10}/\alpha_{em})\nonumber\\
&&=(3.78, -5.97,1.60,-9.94,-1.95,20.9,-175,80.6)\times 10^{-2},
\end{eqnarray}
obtained in Ref.\cite{review} in the VH scheme with 
$\alpha_s = 0.119$, $m_t(m_t) =167$ GeV, $m_b(m_b) = 4.4$ GeV, and $m_c(m_c) = 1.4$ GeV at 
$\mu = 1$ GeV. 

When the anomalous
couplings are included, the values for the WCs $y_i$ 
will change. We will use the leading order QCD
results for $y^{AC}_i$ from the anomalous couplings which are sufficient for the purpose of 
illustrating how the anomalous couplings affect the SM prediction. 
At $\mu = 1$ GeV, anomalous couplings will generate non-zero $y^{AC}_{7,8,9,10}$ as well as 
$y^{AC}_{3,4,5,6}$. We find that the contributions to 
$\epsilon'/\epsilon$ from $y^{AC}_{3,4,5,6}$ are negligibly
small. We will only write down the values for $y^{AC}_{7,8,9,10}$. 
Numerically we have

\begin{eqnarray}
y^{AC}_7/\alpha_{em}&=&-0.243\Delta \kappa^\gamma -0.039\lambda^\gamma +1.21g_1^Z -0.13 g_5^Z,\nonumber\\
y^{AC}_8/\alpha_{em}&=&-0.203\Delta \kappa^\gamma -0.033\lambda^\gamma +0.98g_1^Z -0.11 g_5^Z,\nonumber\\
y^{AC}_9/\alpha_{em}&=&-0.353\Delta \kappa^\gamma -0.057\lambda^\gamma -5.95g_1^Z +0.64 g_5^Z,\nonumber\\
y^{AC}_{10}/\alpha_{em}&=&0.140\Delta \kappa^\gamma +0.023\lambda^\gamma +2.34g_1^Z -0.25 g_5^Z.
\end{eqnarray}
In the above we have used a cut-off $\Lambda = 1$ TeV for terms proportional to $\Delta \kappa^\gamma$ 
and $g_1^Z$.
The above confirms the calculation in Ref.\cite{9}.

Theoretical analyses for $\epsilon'/\epsilon$ are conventionally carried out in terms of the
isospin amplitudes $A_I$ for $K \to \pi\pi$. 
Expressing $\epsilon'/\epsilon$ in terms of $A_I$, 
$y_i$, KM factor $Im(V_{td}V_{ts}^*)$ and hadronic
matrix elements $<O_i>_I = <\pi\pi|O_i|K>_I$, we have

\begin{eqnarray}
{\epsilon'\over \epsilon} &=& 
e^{i(\pi/2+\delta_2-\delta_0)} {\omega\over \sqrt{2} \epsilon}
\left ( {Im A_2\over Re A_2} - {Im A_0\over Re A_0}\right )\nonumber\\ 
&=&e^{i(\pi/2+\delta_2-\delta_0-\phi_\epsilon)}
{G_F \omega\over 2|\epsilon| Re A_0} Im(V_{td}V_{ts}^*)
(\Pi^{SM}_0-{1\over \omega}\Pi^{SM}_2)(1+\Delta)\nonumber\\
\Delta&=& {\Pi^{AC}_0-\Pi^{AC}_2/\omega\over \Pi^{SM}_0-\Pi^{SM}_2/\omega},\nonumber\\
\Pi^{k}_0&=&{1\over \cos\delta_0}\sum_i y^{k}_iRe<O_i>_0 (1-\Omega_{\eta+\eta'}),\;\;
\Pi_2^k={1\over \cos\delta_2}\sum_i y^{k}_iRe<O_i>_2,
\end{eqnarray}
where $\delta_{0}=34.2^0\pm 2.2^0$, $\delta_2 = -6.9^0\pm 0.2^0$\cite{13}
are the final state interaction phases of the amplitudes $A_{0,2}$, and 
$\Omega_{\eta+\eta'}=0.25\pm 0.10$ \cite{10,14}
is correction due to isospin breaking mixing between pion and etas.
$\Delta$ is a measure of the contribution from anomalous couplings
with respect to that from the SM.

To obtain the prediction for $\epsilon'/\epsilon$, one needs to evaluate the hadronic
matrix elements $< O_i>_{0,2}$. This is the most difficult part of the calculation which we will
not attempt to do here. We will take the values listed in Table 6 of 
Ref.\cite{review} which are obtained in Ref.\cite{15}
 using chiral quark model. 
In our numerical calculation, we use: $\delta_0 = 34.2^0$, $\delta_2 = -6.9^0$,
$\omega = 1/22.2$, $\Omega_{\eta+\eta'} = 0.25$.
We have for the SM prediction

\begin{eqnarray}
Re\left ({\epsilon'\over \epsilon}\right )
 \approx 12 Im (V_{td}V_{ts}^*) \approx
12\eta |V_{us}||V_{cb}|^2,
\end{eqnarray}
where $\eta$ is the CP violating parameter in the Wolfenstein parameterization.
We find that the dominant contributions are from terms proportional to $y^{SM}_6$
and $y_8^{SM}$ with the one from $y_8^{SM}$ cancels 
part of the one from $y_6^{SM}$. 

The magnitude of $\epsilon'/\epsilon$
depends on the hadronic parameters mentioned before and also on the KM factor 
$Im(V_{td}V_{ts}^*)\approx \eta |V_{us}||V_{cb}|^2$ 
which is constrained from various experimental measurements.
We use:
$|V_{us}| = 0.2196 \pm 0.0023$, $|V_{cb}| = 0.0395\pm 0.0017$\cite{4}, and
$\eta = 0.381^{+0.061}_{-0.058}$ obtained in Ref.\cite{16}. We note that 
the sign for $\epsilon'/\epsilon$
is positive in agreement with experiment.    
$\epsilon'/\epsilon$ is predicted to be

\begin{eqnarray}
Re\left ({\epsilon'\over \epsilon}\right ) = 1.57^{+0.35}_{-0.34}\times 10^{-3}.
\end{eqnarray}
Here only errors due to KM elements and $\Omega_{\eta+\eta'}$ are 
included.  This range 
is lower than the central experimental data 
 which is a potential problem for the
SM. However we note that even if we take the above theoretical value seriously, 
there is an overlap between the predicted value and the grand average value 
$Re(\epsilon'/\epsilon) = (2.18\pm 0.30)\times 10^{-3}$ of NA31, 
E731 and KTev data at $1\sigma$ level. 
When possible errors\cite{1,15,17} due to 
various parameters such as s quark mass and the 
bag factors $B_i$ are taken into account, the predicted value for $\epsilon'/\epsilon$ 
can be larger\cite{review}. 
It is too early to call for new physics. 
Nevertheless, contributions from 
new physics can dramatically change the SM prediction\cite{18}. 
We now present our analysis for the contributions from anomalous couplings. 
We find indeed that the contributions from anomalous couplings  can enhance
$\epsilon'/\epsilon$ and easy the above potential problem in the SM.

The contributions to $\epsilon'/\epsilon$ due to 
anomalous couplings 
from $I=0$ amplitude are small, 
but contributions from $I = 2$ amplitude through possible large
value for $y_8^{AC}$ can be significantly larger. 
This can be seen by comparing the values of $y^{AC}_8$ 
with the SM value $y_8^{SM} = 0.209\alpha_{em}$.
Using the same hadronic matrix elements used for the SM calculation, 
we obtain

\begin{eqnarray}
Re \left ({\epsilon'\over \epsilon}\right ) \approx 12 Im(V_{td}V_{ts}^*)
(1+0.69\Delta \kappa^\gamma +0.11\lambda^\gamma -2.88g_1^Z +0.31g_5^Z).
\end{eqnarray}

From the above equation we see that
the contributions from anomalous couplings can be large and can also have
the same sign as in the SM, 
in which case the predicted value for $\epsilon'/\epsilon$ 
is enhanced. 
$\Delta \kappa^\gamma <0$, $\lambda^\gamma <0$, $g_1^Z >0$ and 
$g_5^Z<0$ reduce the contribution total value of $\epsilon'/\epsilon$. 
They are therefore not favored.
With the opposite signs, the total contribution to $\epsilon'/\epsilon$ is
enhanced.

It is interesting to note that even in the presence of the anomalous couplings
the constraints on the KM elements do not change, because the anomalous couplings 
do not contribute to the processes used for the fitting. 
We can use the same range 
for the KM factor $Im(V_{td}V_{ts}^*)$. The magnitudes of the contributions to 
$\epsilon'/\epsilon$ depend on the sizes of the anomalous couplings. 
It is easily seen that 
$\epsilon'/\epsilon$ is most sensitive to $g_1^Z$. 
Phenomenological implications of the anomalous couplings have been studied for
high energy collider physics\cite{19,20}, low 
energy flavor conserving processes\cite{21}, and 
flavor changing rare 
decays\cite{22}. Since $K\to \pi\pi$ decays are flavor changing decays, 
we should
use constraints from flavor changing processes for direct comparison.
The conservative allowed ranges at 95\% C.L, with only one anomalous coupling to be 
non-zero in a given process, contains the following ranges:
$\Delta \kappa^\gamma: -0.36\sim 0.4$, $\lambda^\gamma: -1 \sim 1$,
$\Delta g_1^Z: -0.5 \sim 0.1$ and $g_5^Z: -1\sim 2$. Within these ranges,
the anomalous coupling can enhance the contribution to $\epsilon'/\epsilon$
by a factor as large as 2.5
from $g_1^Z$ contribution
which can significantly easy the potential problem in the Standard 
Model. If several anomalous couplings simultaneously contribute with the right 
combination, even larger contribution to $\epsilon'/\epsilon$ can be obtained.
Taking the theoretical calculations for the hadronic matrix elements used here
seriously and require that the prediction and data for 
$\epsilon'/\epsilon$ to have overlap, we find that
 $\Delta \kappa^\gamma <0$, $\lambda^\gamma <0$, $g_1^Z>0$ and
$g_5^Z <0$ are approximately ruled out at $1\sigma$ and $2\sigma$ levels 
for the averaged and KTev values for $\epsilon'/\epsilon$, respectively. 

In conclusion we have studied the contributions of anomalous gauge couplings to 
$\epsilon'/\epsilon$. We find that within the allowed ranges, the 
anomalous couplings can enhance the SM prediction for
 $\epsilon'/\epsilon$ by a factor as large as 2.5. 
This enhancement factor can help easy potential problem in the 
SM which predicts a $\epsilon'/\epsilon$ lower than the experimental data from
some model calculations. $\Delta \kappa^\gamma$, $\lambda^\gamma$, $g_5^Z$ less 
than zero and $g_1^Z$ larger than zero are disfavored.

\noindent {\bf Acknowledgments:} 
I thank C.L. Hsueh and J.-Q. Shi for pointing out a numerical error in an earlier
manuscript.
This work is supported in part by grant NSC 88-2112-M-002-041 of the
R.O.C.


\vspace{1cm}


\begin{thebibliography}{99}


\bibitem{review} For a review see, S. Bertolini, M. Fabbrichesi and J. Eeg, hep-ph/9802405.

\bibitem{1} G. Barr et al., NA31 Collaboration, Phys. Lett. {\bf B317}, 233(1993).

\bibitem{2} L. Gibbons et al., E731 Collaboration, Phys. Rev.  {\bf D55}, 6625(1997).

\bibitem{3} KTev Collaboration, 
http://fnphyx-www.fnal.gov/experiments/ktev/ktev.html.

\bibitem{4} Particle Data Group, Eru. Phys. J. {\bf C3}, 1(1998).

\bibitem{5} For reviews see, Xiao-Gang He, e-print hep-ph/9710551, 
CP Violation, in proceedings of
CP Violation and Various Frontiers in Tau and Other systems, CCAST, 
Beijing, China, 11-14 August, 1997; C. Jarskog, CP Violation, World Scientific, Singapore
(1989).

\bibitem{6} L. Wolfenstein, Phys. Rev. Lett. {\bf 13}, 562(1964).

\bibitem{7} M. Kobayashi and T. Maskawa, Prog. Theor. Phys. {\bf 49}, 652(1973).

\bibitem{8} T. D. Lee, Phys. Rev. {\bf D8}, 1226(1973); Phys. Rep. {\bf 96}, 143(1976); 
S. Weinberg, Phys. Rev. Lett. {\bf 31}, 657(1976).

\bibitem{9} X.-G. He and B. McKellar, Phys. Rev. {\bf D51}, 6484(1995).

\bibitem{10} J. Donoghue et al,, Phys. Lett.
{\bf B179}, 361(1986);
J. Flynn and L. Randall, Phys. Lett. {\bf B224}, 22(1989); Erratum {\bf B235}, 412(1990).

\bibitem{11} K. Gaemers and G. Gunaris, Z. Phys. {\bf C1}, 259(1979);
K. Hagiwara et al., Nucl. Phys. {\bf B282}, 253(1987).

\bibitem{12} M. Lusignoli, Nucl. Phys. {\bf B325}, 33(1989);
A. Buras, M. Jamin and M. Lautenbacher, Nucl. Phys. {\bf B408}, 209(1993);
M. Cuichini et al., Nucl. Phys. {\bf B415}, 403(1994).

\bibitem{dhp} S. Bertolini, J. Eeg and M. Fabbrichesi, Nucl. Phys. {\bf B449}, 197(1995);
N. Deshpande, X.-G. He and S. Pakvasa, Phys. Lett. {\bf B326}, 307(1994).

\bibitem{13} E. Chell and M. Olsson, Phys. Rev. {\bf D48}, 4076(1993).

\bibitem{14} A. Buras and J. Gerard, Phys. Lett. {\bf B192}, 156(1987);
H.-Y. Cheng, Phys. Lett. {\bf B201}, 155(1988).


\bibitem{15} S. Bertolini et al., Nucl. Phys. 
{\bf B514}, 93(1998).

\bibitem{16} S. Mele, e-print hep-ph/9810333.

\bibitem{17} A. Buras, M. Jamin and M. Lautenbacher, Phys. Lett. {\bf B389}, 749(1996);
M. Cuichini et al., 
Z. Phys. {\bf C68}, 239(1995); E. Paschos and Y. L. Wu, Mod. Phys. Lett. {\bf A6}, 93(1991).

\bibitem{18} 
J. Donoghue and B. Holstein, Phys. Rev. {\bf D32}, 1152(1985);
H.-Y. Cheng, Phys. Rev. {\bf D34}, 1397(1986);
R. Mohapatra and J. Pati, Phys. Rev. {\bf D11}, 566(1975);
D. Chang et al., Phys. Rev. {\bf D30}, 1601(1984);
X.-G. He, B. McKellar and S. Pakvasa, Phys. Rev. Lett. {\bf 61}, 1267(1988);
H. Koning, Z. Phys. {\bf 73}, 161(1996);
J. Agrawal and P. Frampton, Nucl. Phys. {\bf B419}, 254(1994).

\bibitem{19} The LEP electroweak working group, LEPEWWG/TGC/97-01;
F. Abe et al.,  CDF Collaboration, Phys. Rev. Lett. {\bf 78}, 4536(1997);
S. Abachi et al., D0 Collaboration, Phys. Rev. Lett. {\bf 78}, 3640(1997);

\bibitem{20} C. P. Burgess and D. London, 
Phys. Rev. Lett. {\bf 69}, 3428(1992); 
A. De Rujula et al., Nucl. Phys. {\bf B384}, 3(1992);
K. Hagiwara et al., Phys. Lett. {\bf B283}, 353(1992);
P. Hernandes and F. Vegas, Phys. Lett. {\bf B307}, 116(1993).

\bibitem{21} A. Grau and J. Grifols, Phys. Lett. {\bf B154}, 283(1985);
J. Wallet, Phys. Rev. {\bf D32}, 813(1985).


\bibitem{22} S.-P. Chia, Phys. Lett. {\bf B240}, 465(1990);
K. A. Peterson, Phys. Lett. {\bf B282}, 207(1992);
K. Namuta, Z. Phys. {\bf C52}, 691(1991);
T. Rizzo, Phys. Lett. {\bf B315}, 471(1993);
X.-G. He, Phys. Lett. {\bf B319}, 327(1993);
X.-G. He and B. McKellar, Phys. Lett. {\bf 320}, 168(1994);
S. Dawson and G. Valencia, Phys. Rev. {\bf D49}, 2188(1994);
G. Baillie, Z. Phys. {\bf C61}, 667(1994);
G. Burdman, Phys. Rev. {\bf D59}, 035001(1999).

\end{thebibliography}
\end{document}